\documentclass{article}
\usepackage[utf8]{inputenc}
\usepackage{amssymb}
\usepackage{amsmath}
\usepackage{amsfonts}
\usepackage{mathtools}
\usepackage{bbm}

\usepackage{enumitem}

\usepackage{xcolor}

\usepackage{tensor}
\usepackage{booktabs}
\usepackage{physics}
\usepackage{tensor}
\usepackage[papersize={8.5in,11in}]{geometry}
 
 \newcommand{\ii} {\mathrm{i}}

\title{Vacuum Petrov type D horizons of nontrivial $U(1)$ bundle structure over Riemann surfaces with genus $> 0$}

\author{Jerzy Lewandowski \thanks{Jerzy.Lewandowski@fuw.edu.pl} \and Maciej Ossowski \thanks{Maciej.Ossowski@fuw.edu.pl}}

\date{%
\small
    Faculty of Physics, University of Warsaw,\\
ul. Pasteura 5, 02-093 Warsaw, Poland\\[2ex]%
    \today
}

\begin{document}

\maketitle

\begin{abstract}
{\color{black} We consider isolated horizons (Killing horizons up to the second order) whose null flow has the structure of a U(1) principal fiber bundle over a compact Riemann surface. 
We impose the vacuum Einstein equations (with the cosmological constant) and the condition that the spacetime Weyl tensor is of Petrov D type on the geometry of the horizons. 
We derive all the solutions in the case when the genus of the surface is $>1$. By doing so for all the nontrivial bundles, we complete the classification. }

We construct the embedding spacetimes and show that they are locally isometric to the toroidal or hyperbolic generalization of the Taub-NUT-(anti-) de Sitter spacetimes for horizons of genus $1$ or $>1$ respectively, after performing Misner's identification of the spacetime.
The horizon bundle structure can be naturally extended to bundle structure defined on the entire spacetime.
\end{abstract}

\tableofcontents
\section{Introduction}

Einstein's equations induce equations for the geometry of the Killing horizon.
These horizon equations determine - from within - the properties of horizons in a manner similar to the theorems of the global mathematical theory of black holes.
They imply the topology of global sections (in the rotating case), rigidity, and no-hair \cite{localnohairPhysRevD.98.024008,Lewandowski_2006,Lewandowski_2003}.
Sometimes they even reach where the global theory does not, as is the case, for example, with extreme horizons and their equations \cite{Chruściel2012}.

{\color{black} Isolated horizon theory provides a local framework for the study of the black hole horizons without assumptions about the horizon's neighborhood.
This approach may be used in numerical relativity where the existence of the blackhole defining Killing vector field in the spacetimes is not given.
In mathematical relativity, they have been employed in classification of extreme horizons \cite{PodolskyMatejov22}. 
Analogously to the Killing horizons, the isolated horizons obey the laws of the so-called black hole thermodynamics, as long as the Einstein equations are imposed \cite{MechanicsIHPhysRevD.64.044016}.  
It is worth noting that they naturally appear as leaves of foliations of some classes of spacetimes, for example there exist solutions in Kundt's class foliated by a weaker structure than isolated horizons, i.e. nonexpanding horizons \cite{Pawlowski_2004}.
The famous Robinson-Trautmann solutions admit an isolated horizon, which however is generically not a Killing horizon \cite{Robinson1962}.
Finally, although commonly used to study black hole horizons, a broader notion of a weakly isolated horizon has been recently successfully applied to the null infinity of the spacetime \cite{ahtekarSpeziale2024,ahtekar2024}.
}

In this paper, we continue the study of the equation implied by the assumption that the nonextreme horizon in $\Lambda$-vacuum 4-dimensional spacetime is of the Petrov type D. 
In an earlier study, we investigated the Petrov type D equation on horizons that allowed for a global spacelike section that was a compact 2-dimensional surface \cite{localnohairPhysRevD.98.024008,TypeDGenus,typeD}.
Our equation takes the form of the vanishing second holomorphic derivative of a complex function formed from the scalar of the curvature of the surface and the rotation represented by an exact 2-form. 
A certain subtlety in this case is the nontrivial first group of cohomology admitting nonequivalent closed 1-form rotation potentials: 
different elements of the first group of cohomology define different horizons.
If the Euler characteristic of the surface is nonpositive, the only solutions are 2-geometries of constant curvature and the rotation 2-form vanishing at every point.
So, in this class, rotating solutions can only exist when the section of the horizon is a sphere. 
Then, the family of axisymmetric solutions is parameterized by three numbers: area, angular momentum, and cosmological constant. 
They are in $1-1$ correspondence with the Kerr - (anti-) de Sitter spacetimes. 
Moreover, the necessity of the axial symmetry can be proved by assuming that the horizon of Petrov type D is bifurcated. 
All those solutions, horizons admitting global sections, are also embeddable in known solutions of Einstein's equations. 
    
A horizon that admits a global section has the structure of a trivial principal fiber bundle with the null generators as the fibers.  
The next class of horizons we consider, are such that the null flow defines the structure of a nontrivial $U(1)$ principal fiber bundle over a compact manifold. 
We have already studied in earlier papers the case of the Hopf bundle and the bundles with higher topological charges over a (topological) 2-sphere \cite{LO,LO2,LO3}.  
The unknowns of the vacuum type D equation become the metric induced on the sphere and the curvature of a connection defined on the bundle (multiplied by any nonzero surface gravity).  
All axisymmetric solutions were found. 
They form a family parametrized by four parameters: the surface area, the angular momentum (or the Kerr parameter), the cosmological constant, and the NUT parameter, and are embeddable in the Kerr- NUT - (Anti-) de Sitter spacetimes or their accelerated generalization.
Recently horizons with a structure of a Hopf bundle whose action is transversal to the null direction have been introduced and studied \cite{LDROTransversal}.
Such horizons arise naturally in Kerr-NUT-(anti-) de Sitter spacetimes where orbits of nonnull direction have been compactified to $U(1)$ to achieve a regular axis of rotation \cite{LO3}.
After performing the procedure the space of null generators may still have conical singularities.
Petrov type D equation on a sphere admitting conical singularities has been solved and the solutions are found to be embeddable in accelerated Kerr-NUT-(anti-)de Sitter spacetimes with period time coordinate identification.

In the present work, we turn to the study of type D horizons, whose null Killing flow defines a nontrivial $U(1)$ principal fiber bundle over a 2-dimensional with a nonzero genus.  
Also, we construct NUT-type spacetimes containing our solutions.  
In this paper we consider all objects to be smooth. 
It is possible to weaken this assumption up to the fourth order of differentiability, but not lower, since the type D equation of this order has to be imposed.

%\section{Nonembedded isolated horizons and the vacuum Petrov type D equation.}
\subsection{Nonembedded isolated horizons}
The general theory of isolated horizons has been widely studied in literature.
Based on \cite{Ashtekar2004,geometryhorizonsAshtekar_2002,AshtekarDynamicalhorizonsandtheirproperties,typeD}
we recall the definition and properties important for the current paper.  
In this subsection we present an entirely local approach, keeping any global properties for later assumptions.   

Consider a manifold $H$, which will serve as a model for the horizons, endowed with $(i)$ a degenerate metric tensor $g$ of the signature $(0+\ldots+)$,  and  $(ii)$ a torsion-free covariant derivative $\nabla$, that preserves the degenerate metric tensor 
\begin{equation*}
    \nabla g =0.
\end{equation*}
The triple $(H,g,\nabla)$ is called a {\it nonexpanding horizon}.
The degeneracy of $g$ has two consequences; 
One of them is that $\nabla$ cannot be uniquely determined by $g$. The second is that $\nabla$ may not exist at all unless it is true that any null vector field $\ell$, that is such that
 \begin{equation*}
    g(\ell,\cdot)=0,
\end{equation*}
generates a symmetry of $g$, namely it satisfies
\begin{equation*}
    \mathcal{L}_\ell g = 0.
\end{equation*}
This second property justifies the name "nonexpanding".
% and and legitimizes even: shear-free.  {\color{red} ?}
Conversely, given $g$ the latter condition is sufficient for a family of the covariant derivatives $\nabla$ to exist.    
In fact, if any nonvanishing null vector field $\ell$ generates locally a symmetry of $g$, then so does $\ell'=f\ell$ for any function $f{\in C^\infty(H)}$. 

{\it Isolated horizon} is 4-tuple: $(H,g,\nabla,[\ell])$,  where $(H,g,\nabla)$ is a nonexpanding horizon, and  $\ell$ is a nowhere vanishing null vector field  defined up to rescaling by a constant factor 
\begin{equation}\label{resc}
  \ell\mapsto a_0 \ell, \quad  a_0\in \mathbb{R}
\end{equation}
such that
  \begin{equation}
\label{eq:stationary-to-second-order}
    [\mathcal{L}_\ell,\nabla]=0,
\end{equation}
 {\it The rotation $1$-form potential}   of a nonexpanding  horizon is a $1$-form $\omega$  defined by a null vector field $\ell$ as follows, 
 \begin{equation*}
    \nabla_a \ell^b = \omega_a \ell^b.
\end{equation*} 
It exists due to the fact, that the null direction has to be covariantly constant due to the metricity of $\nabla$. 
In the case of an isolated horizon 
% $(H,g,\nabla,\ell)$  
$(H,g,\nabla,[\ell])$, the rotation $1$-form potential is uniquely defined and satisfies
\begin{equation}
    {\cal L}_\ell \omega = 0. 
\end{equation}
The surface gravity $\kappa$ is defined as the self-acceleration of $\ell$
\begin{equation}
    \nabla_\ell \ell = \kappa \ell
\end{equation}
and is defined up to (\ref{resc}), upon which it transforms as:
\begin{equation}
    \kappa \mapsto a_0\kappa.
\end{equation}
Clearly we have $\ell(\kappa)=0$. 
If the isolated horizon were embedded in spacetime, then assumptions about the spacetime Riemann tensor would imply various constraints on $g$ and $\nabla$, which can be just introduced for the nonembedded isolated horizon as well.
One of them is the $0-$th law of isolated horizon thermodynamics, namely
\begin{equation}\label{0th}
    \kappa = \rm const,
\end{equation}
on every connected component of $H$.  
Indeed, in the embedded case it would be implied by Einstein's equations and energy inequalities.  
Whenever
\begin{equation}
    \kappa\not=0,
\end{equation}
we call an isolated horizon {\it nonextremal} and vice versa.  

Remaining components of $\nabla$ can be determined by the action of $\nabla$ on a given $1$-form $n$ on $H$, such that
$n_a\ell^a=-1$.

\subsection{Isolated horizon of the principal fiber bundle structure and null fibers}
Consider an isolated horizon $(H,g,\nabla,[\ell])$. 
Suppose the null flow of the vector field $\ell$ defines globally the action of a group $G=\mathbb{R}$ or $G=U(1)$, that provides $H$ with a structure of a principal fiber bundle over a space $S$ of the null curves in $H$, and the canonical projection
\begin{equation}
\Pi: H\rightarrow S.  
\end{equation}
Then, the degenerate metric tensor $g$ defined on $H$ induces a Riemannian metric tensor $g^S$ on $S$ such that 
\begin{equation}
g = \Pi^* g^S. 
\end{equation}

The bundle structure is nontrivial only in the case $G=U(1)$, and then we can fix the normalization of $\ell$ such that it is consistent with the parametrization of $U(1)$ by numbers from the interval $[0,2\pi)$. 
That makes the surface gravity $\kappa$ also a uniquely defined quantity. 

The role of the rotation potential $1$-form $\omega$ in the bundle structure depends on whether $H$ is extremal or nonextremal. 
In the case of a nonextremal horizon, $\omega$ gives rise to a $G$-connection
\begin{equation}\label{G-connection}
A := \frac{1}{\kappa}\omega \otimes \ell^*,
\end{equation}
where $\ell^*\in \mathbb{R}$ or $\ell^*\in u(1)$ (the Lie algebra of $G$) corresponds to the vector field $\ell$.  
Similarly, the $G$-curvature reads
\begin{equation}
    F:=\frac{1}{\kappa}\dd\omega \otimes \ell^*.
\end{equation}
As the the Lie algebra is one dimensional all information about the connection and curvature is already contained in the 1-form $\frac{1}{\kappa}\omega$ and its derivative.
Therefore when referring to the connection and curvature we will omit the Lie algebra valued part, bearing in mind that now the normalization condition $A(\ell)=1$ is required.

In the consequence, the rotation $1$-form may be pulled back to $S$ by a local section
\begin{equation}
\omega^\sigma = \sigma^*\omega.    
\end{equation}
The pullback is however section-dependent, and if the bundle is nontrivial, a family of different sections has to be used to cover all the $S$. 
Still, the {\it rotation $2$-form} $\dd\omega$
provides a uniquely and globally defined $2$-form on $S$ that can be constructed using local sections 
\begin{equation}
 \tilde{\Omega}^S = \sigma^* d\omega .    
\end{equation}
If the horizon $H$ is extremal, then the pullback  
\begin{equation}
\omega^S := \sigma^*\omega    
\end{equation}
is independent of $\sigma$ and the $1$-form $\omega^S$ is globally defined on $S$. 
In that case, $g^S$ and $\omega^S$ are insensitive to the nontrivial bundle structure and the results of  \cite{TypeDGenus} apply. Therefore in this paper, we assume that the horizons are necessarily nonextremal.  

A structure that we do not consider in this paper, is an isolated horizon with a principal fiber bundle structure whose fibers are transversal to the null direction.
For the discussion of such horizons with $S$ being a topological two sphere see \cite{LDROTransversal}.

\subsection{$3$-dimensional isolated horizons and the type D equation}
On a $3$-dimensional isolated horizon, we define a {\it curvature scalar} $K$ and {\it rotation pseudoscalar} $\Omega$.
To introduce them at a point $x\in H$, consider any $2$-submanifold $S'\subset H$ transversal to $\ell$  at $x$.
The curvature scalar $K(x)$ at  $x$ is the Gauss curvature of the  Riemannian geometry induced on $S'$ by the degenerate metric tensor $g$ of $H$, and its value at $x$ is independent of the choice of $S'$ at $x$.
The {\it rotation pseudo-scalar} $\Omega(x)$ is given by the  pullback  $d\omega^{S'}$ of $d\omega$ to $S'$ and the $2$-area form   $\eta^{S'}$ of $S'$.
It is defined as
\begin{equation}
\label{eq:def-Omega}
    d\omega^{S'} =:\Omega\ \eta^{S'}.
\end{equation}
The value $\Omega(x)$  depends on the orientation of $S'$, therefore in a neighborhood of $x$, we orient all the $ 2$ surfaces consistently. 
To define $\Omega$ globally on $H$ we need a globally defined orientation of the $2$-dimensional sections. 

Both of the functions $K$ and $\Omega$ are constant along the null curves
\begin{equation}
    \ell(K)=\ell(\Omega) =0. 
\end{equation}
We are assuming throughout this paper that the $0$th law (\ref{0th}) holds. 

The functions $K$ and $\Omega$ set a complex valued function $K+\ii\Omega-\frac{1}{3}\Lambda$ depending on the parameter $\Lambda$ called  {\it cosmological constant}.
On this function the {\it vacuum Petrov type D equation} is imposed.
To write it out explicitly, consider again an arbitrary $2$-dimensional submanifold $S'\subset H$ transversal to the null vector field $\ell$.
Introduce on it a null complex coframe $m^{S'}_A$ ($A,B \in 1,2$) such that the induced metric tensor and the area $2$-form read
\begin{equation}
    g^{S'}=m^{S'}_A \bar{m}^{S'}_B + m^{S'}_B \bar{m}^{S'}_A,\quad  \eta^{S'}=\ii( \bar{m}^{S'}_A m^{S'}_B - \bar{m}^{S'}_B m^{S'}_A).
\end{equation}
Then the type D equation is defined as follows \cite{typeD}: 
\begin{equation}\label{theEq}
    \bar{m}^{S'A}\bar{m}^{S'B} \nabla^{S'}_A \nabla^{S'}_B \left(K+\ii \Omega-\frac{\Lambda}{3}\right)^{-1/3}=0,
\end{equation}
where $\nabla^{S'}_A$ is the torsion-free and metric covariant derivative determined on the $2$ manifold by $g^{S'}_{AB}$,  and the functions $K$ and $\Omega$ are assumed to satisfy
\begin{equation}
\label{eq:fPetrov-nonvanishing}
    K+\ii \Omega-\frac{\Lambda}{3} \neq 0.
\end{equation}

% {\color{red} Ten index S' myli sie z indeksami A,B, warto pomyslec o zmianie oznaczen}

This equation stems from the theory of embedded isolated horizons. 
It is a necessary condition for a $3$ dimensional nonextremal isolated horizon $H$ to be embeddable in a $4$ dimensional spacetime such that: 
 \begin{enumerate}[label=(\roman*)]
\item The vacuum Einstein equations with cosmological constant $\Lambda$ are satisfied at $H$,
\item The Weyl tensor is of the Petrov type D at $H$, and
\item The Weyl tensor is constant along $H$ with respect to a suitable extension of the vector field $\ell$ (see \cite{typeD} for the details). 
\end{enumerate}

\noindent{\bf Remark} Suppose an isolated horizon $H$  of the degenerate metric $g$ and the rotation $1$-form potential $\omega$ is a solution to the vacuum Petrov type D equation (\ref{theEq}) with a cosmological constant $\Lambda$. Then so is $H$ endowed with any
\begin{equation}
    g' = \alpha^2 g, \ \ \ \ \omega'=\omega, \ \ \ \ \Lambda'=\frac{\Lambda}{\alpha^2}, \ \ \ \ \alpha\in \mathbb{R}\setminus\{0\}. 
\end{equation}
An example of a solution is given by $g$ and $\omega$ such that 
\begin{equation}
K = {\rm const}, \ \ \ \ \Omega = {\rm const}    
\end{equation}
such that (\ref{eq:fPetrov-nonvanishing}) for a proper type D. 
Then, an even stronger statement is true: every
\begin{equation}
    g''=\alpha^2g, \ \ \ \ \omega''=\beta^2\omega, \ \ \ \ \alpha, \beta \in \mathbb{R}\setminus\{0\}
\end{equation}
is another solution to (\ref{theEq}) with arbitrary cosmological constant
\begin{equation}
 \Lambda''=\gamma\Lambda, \ \ \ \ \gamma\in \mathbb{R}
\end{equation}
where again (\ref{eq:fPetrov-nonvanishing}) is assumed. 
% }
% {\color{red} cos nie gra z a i b ? inne oznaczenia? Wszystko gra, tylko ma miejsce pewna gradacja. W ogolnym przypadku mamy te symetrie w przestrzeni rozwiazan ze stala a. Ale gdy K oraz $\Omega$ sa const, symetrii jest wiecej, pojawiaja sie jeszcze stala b oraz c.}

Suppose, finally, that a $3$ dimensional horizon $H$ has the principal fiber bundle structure of null fibers considered above.
Then the space $S$ of the null generators is endowed with the Riemannian metric tensor $g^S$, the curvature scalar  $K$, and the rotation pseudoscalar $\Omega$.
Now $K$ becomes the Gauss curvature of $g^S$.
Moreover, if $S$ is compact, then 
\begin{equation}
\label{Euler}
    \int_S K\eta^S = 2\pi \chi_E(S),
\end{equation}
where $\chi_E$ is the Euler characteristics of $(S,g^S)$ while
\begin{equation}
\label{Chern}
    \int_S \Omega\eta^S = 2\pi \kappa \chi_C(H)_,
\end{equation}
 where $\chi_C$ is the Chern number of the bundle $H$, and $\kappa$ is the surface gravity introduced above.  The Petrov type D equation amounts now, to the equation
 \begin{equation}\label{theeq}
    \bar{m}^{SA}\bar{m}^{SB} \nabla^{S}_A \nabla^{S}_B \left(K+\ii \Omega-\frac{\Lambda}{3}\right)^{-1/3}=0,
\end{equation}
where the unknowns are  the metric tensor $g^S$ and the family of the  $1$-forms $\omega^\sigma$, i.e. the local pullbacks of the rotation $1$-form $\omega$.    

In the case of a trivial bundle $H$ over a topological sphere $S=S^2$ and axially symmetric $(g^S,\omega^S)$, the Petrov type D equation is satisfied if and only if $g^S$ and $\omega^S$  correspond (via an embedding in  $4$ dimensional spacetime) with those induced on the space of the null generators of a Killing horizon in either the Kerr, Kerr-de Sitter, or Kerr-anti-de Sitter spacetime, depending on the sign of $\Lambda$ \cite{localnohairPhysRevD.98.024008}.  
%If the Killing horizon is nonextremal,  then  covariant derivative induced on it  agrees with %$\nabla$ defined on the  horizon $H$. In the extremal Killing horizon case, though, the induced   %or in the so called Near Horizon Geometry spacetimes \cite{localnohairPhysRevD.98.024008}.
The axial symmetry can be derived if we assume that the horizon is bifurcated (amounting to two horizons attached to each other at the asymptotic boundaries) \cite{PhysRevD.97.124067}. 
The existence of nonaxisymmetric solutions is an open problem.

For nontrivial bundles over the topological sphere, again all the axisymmetric solutions of (\ref{theeq}) were found \cite{hopf}, some (whether all is an open problem) are embeddable in the Kerr-NUT-(anti) de Sitter spacetimes \cite{hopf,LO2,LDROTransversal}.

For the bundles over compact 2-manifolds of nonpositive Euler characteristics, the only solutions to the Petrov type D equation (\ref{theeq}) are those, that satisfy \cite{TypeDGenus}
 \begin{equation}
 \label{KOmegaconst}
K= {\rm const}, \ \ \ \ {\rm and} \ \ \ \ \Omega = {\rm const} .    
 \end{equation}
%In the trivial budle case, that implies
%\begin{equation}
 %   \Omega = 0.
%\end{equation}
This case is the focus of this paper. 
We construct all the vacuum Petrov type D isolated horizons in this class. 
Next, we find globally defined spacetimes, exact solutions to Einstein's vacuum equations, that accommodate them upon a suitable embedding. 

\section{Vacuum Petrov type D isolated horizons whose null flow \\ has PFB structure over a Riemann surface of genus $g\ge 1$}
\subsection{The horizons over $2$-torus}
\label{sec:torus-horizons}
Consider now the case, when the space of the null directions $S$ of the vacuum type D isolated horizon $H$ is a 2-dimensional torus   
\begin{equation}
   S = S_1\times S_1 . 
\end{equation}
The first of the  Eqs. (\ref{KOmegaconst}) combined with the Eq. (\ref{Euler}) imply 
\begin{equation}
    K = 0,
\end{equation}
hence the metric  $\prescript{(2)}{}{g}$  has to be flat. 
Therefore, there exist cyclic coordinates $(\phi,\psi)\in [0,2\pi)\times[0,2\pi)$ that set a coordinate system on $S_1\times S_1$,  such that  {
\begin{equation}\label{gtorus}
    {g^S}=\frac{1}{P^2_0}(a^2 \dd \phi^2+2ab\ \dd \psi \dd \phi+ (1+b^2)\dd \psi^2),
\end{equation}
with real constants $a>0,\ b$ and $ P_0>0$.   
% We will also use the area $2$-form, it is
%\begin{equation}
 %   \prescript{(2)}{}\eta=\frac{a}{P_0^2}\dd\psi \wedge \dd \phi.
%\end{equation}
The above admits the following discrete symmetries 
\begin{enumerate}
    \item If $b\not=0$, then we can assume $b>0$ without loss of generality,  due to either $\psi\mapsto-\psi$ or $\phi\mapsto-\phi$  and the remaining coordinate freedom is a simultaneous change $(\psi,\phi)\mapsto (-\psi,-\phi)$.  
    \item If $b=0$, then the remaining coordinate freedom is independently  $\psi\mapsto -\psi$ and / or $\phi\mapsto -\phi$. 
\end{enumerate}

The rotation $1$-form potential depends on the bundle structure.
Therefore we will split our considerations into the trivial and nontrivial cases.

\subsubsection{The trivial bundle case}
\label{sec:trivial-torus-horizons}
In the case of a horizon of the structure of a trivial PFB, Eq (\ref{Chern}) and the second equality in (\ref{KOmegaconst}) imply 
\begin{equation}\label{Omega0}
    \Omega = 0.
\end{equation}
Therefore, a corresponding $G$-connection $A$ (\ref{G-connection}) is flat, and it can be represented by a single closed $1$-form on $S$, 

\begin{equation}
 A^S = \alpha \dd\phi + \beta \dd\psi + \dd f, \ \ \ \alpha,\beta \in \mathbb{R},\ f\in C^\infty(S_1\times S_2) ,  
\end{equation}
defined modulo global gauge transformations (see below). 

The horizon over $S$ has the topology
\begin{equation}
    H = G\times S 
\end{equation}
where $G=\mathbb{R}, \ U(1)$. 
Let $\tau$ be a coordinate along $G$ with $(\tau,\psi,\phi)$ being the corresponding coordinate system on $H$ such that the null vector field $\ell$  (modulo a constant, however see below) is
 \begin{equation}
     \ell = \partial_\tau .
 \end{equation}
Then, the degenerate metric tensor of $H$ is given by just the same formula as above, namely
 \begin{equation}\label{gH-torus}
    {g}=\frac{1}{P^2_0}(a^2 \dd \phi^2+2ab\ \dd \psi \dd \phi+ (1+b^2)\dd \psi^2).
\end{equation}
The corresponding $G$-connection on $H$ viewed as the PFB is
\begin{equation}
 A = \left( \dd\tau + \alpha \dd\phi + \beta \dd\psi + \dd f\right) \otimes \partial_\tau,  
\end{equation}
while the formula for the rotation $1$-form potential such that (\ref{Omega0}) reads
\begin{equation}
% \label{omega-triv}
 \omega = \kappa \left( \dd\tau + \alpha \dd\phi + \beta \dd\psi + \dd f\right).
\end{equation} 
  
The term $df$ can be eliminated by a change of the coordinates
$$\tau= \tau' + f,$$
\begin{equation}
% \label{omega-triv'}
 \omega = \kappa \left( \dd\tau' + \alpha \dd\phi + \beta \dd\psi\right).
 \end{equation} 

Now, if the gauge group $G=\mathbb{R}$ then, as explained above, rescaling freedom of the vector field $\ell$ (or, in other words, the coordinate $\tau$) allows us to set $\kappa=1$. 
The coefficients $\alpha$ and $\beta$ may be made nonnegative $\alpha,\beta\ge 0$, however, their range cannot be further reduced, and hence they are true degrees of freedom.
 
If the structure group of the PFB of $H$ is $G=U(1)$, then the scaling ambiguity of the vector field $\ell$ can be fixed, by choosing $\tau$ to range $[0,2\pi)$. 
Therefore, in that case, the value of the surface gravity $\kappa$ is an intrinsically meaningful characteristic of the horizon. 
On the other hand, there are more gauge transformations, meaning bundle automorphisms, namely
$$\tau" = \tau' + n\psi + m\phi, \ \ \ \ m,n\in \mathbb{Z}. $$
Hence, in the $U(1)$ case,  without loss of generality, it is sufficient to consider $\alpha,\beta\in [0,1)$. 

In conclusion,  the meaningful degrees of freedom of $g$ and $\omega$ split into the following cases
\begin{enumerate}
\item If $b=0$ and  $G=\mathbb{R}$, then $\kappa\in\mathbb{R}\setminus \{0\}$, and $\alpha,\beta \ge 0$.
\item If $b>0$ and  $G=\mathbb{R}$, then $\kappa\in\mathbb{R}\setminus \{0\}$, and $\alpha,\beta \in \mathbb{R}$, which are determined up to  $(\alpha,\beta)\mapsto (-\alpha,-\beta)$ 
\item If $G=U(1)$, then $\kappa\in \mathbb{R}^+, \ \alpha,\beta \in [0,1)$, modulo $\alpha\mapsto 1-\alpha$ and / or  $\beta\mapsto 1-\beta$, in the $b=0$ case, and modulo  $(\alpha, \beta)\mapsto (1-\alpha,  1-\beta)$, in the $b>0$ case. 
\end{enumerate}

 \noindent
 {\color{black}
The corresponding covariant derivative $\nabla$ is determined on $H$ by the metric connection of $g^S$ and the rotation 1-form $\omega$ \cite{typeD}. 
}

\subsubsection{nontrivial bundle case} 
Consider now a nontrivial $U(1)$-bundle over $S:=S^1\times S^1$ (there are no nontrivial $\mathbb{R}-$bundles, due to $\mathbb{R}$ being contractible). 
It may be characterized using an open cover $\{V_1, V_2\}$ of $S^1\times S^1$ and a  transition function $a_{12}: V_1\cap V_2\to U(1)$ defined modulo 
\begin{equation}\label{gauge}
a_{12}\mapsto a'_{12} = b_1^{-1}a_{12}b_2, \ \ \ b_i:V_i\rightarrow U(1), i=1,2.      
\end{equation}
It is sufficient to find just one pair $(\tilde{A}^S_{(1)}, \ \tilde{A}^S_{(2)})$, and every other $(A^S_{(1)}, \ A^S_{(2)})$ will be given by a globally defined $1$-form $a$ on $S^1\times S^1$, namely
\begin{equation}\label{AS}
 A^S_{(i)} = \tilde{A}^S_{(i)} + a.    
\end{equation}
The connection  $A^S_{(i)}$ will be used to solve the second equality of (\ref{KOmegaconst}) provided its Hodge dual is a constant function, 
\begin{equation} 
\label{eq:hodge-star-A}
 *\dd A^S_{(i)} = \rm const,   
\end{equation}
where the area $2$-form $\eta$ of the metric tensor (\ref{gtorus}) is
\begin{equation}
    \prescript{(2)}{}\eta=\frac{a}{ P_0^2}\dd\psi \wedge \dd \phi.
\end{equation}
{\color{black} The existence of $A^S$ satisfying (\ref{eq:hodge-star-A}) is demonstrated explicitly for $S$ being a torus using coordinates. The existence is also obvious for a U(1) bundle automorphic to the bundle of the orthonormal frames.  For higher genus surfaces we take the existence as an assumption.}

A possible choice of the open cover is (we identify below $S^1$ with the interval $[0,2\pi)$ endowed with the compact topology)
\begin{equation*}
\begin{split}
    &V_1=\big(\epsilon\ ;\ 2\pi-\epsilon\big]\times S^1,\quad V_2=\big[ 0\ ;\ \pi - \epsilon \big)\ \cup \ \big (\pi+\epsilon \ ;\ 2 \pi\big]\times S^1,\\
    &V_1\cap V_2=\big(\epsilon\ ;\ \pi-\epsilon \big)\times S^1 \ \cup\ \big(\pi+\epsilon\ ;\ 2\pi-\epsilon \big)\times S^1,
\end{split}
\end{equation*}
for $0<\epsilon<\tfrac{\pi}{2}$. 

Using the gauge freedom (\ref{gauge}) we can set the transition function to
\begin{equation}
\label{eq:gauge-U(1)}
    a_{12}(\psi,\phi):=
    \begin{cases}
    1 & \text{at } \big(\epsilon\ ;\ \pi-\epsilon \big)\times S^1\\
    \exp(-\ii \chi_C \phi) & \text{at }\big(\pi+\epsilon\ ;\ 2\pi-\epsilon \big)\times S^1, \ \ \chi_C\in \mathbb{Z}.
    \end{cases}
\end{equation}
Next step is to construct the local connection 1-forms $\tilde{A}^S_{(i)}$, each defined in $V_i$.  
Consider a 1-form 
\begin{equation}
    \tilde{A}^S_{(1)}=\chi_C \frac{\psi}{2\pi}\dd\phi.
\end{equation}
Indeed, it is well defined in $V_1$, as $\psi$ does not have a discontinuity there. 
Next, we define on  $V_1\cap V_2$  
\begin{equation}
\label{eq:gauge-transform}
    \tilde{A}^S_{(2)}= \tilde{A}^S_{(1)}-\ii a_{12}^{-1}\dd a_{12},
\end{equation} 
and extend it by the analyticity to $V_2$, 
hence we obtain
\begin{equation}
   \tilde{A}^S_{(2)}=
        \begin{cases}
    \chi_C \frac{\psi}{2\pi}\dd\phi & \text{at } \big[0\ ;\ \pi-\epsilon \big)\times S^1,\\
    \chi_C \frac{\psi-2\pi}{2\pi}\dd\phi & \text{at }\big(\pi+\epsilon\ ;\ 2\pi\big]\times S^1.
    \end{cases}
\end{equation}
To restore symmetry between the two elements of the covering,  consider a function $\psi'$ continuous on $V_2$ but with a step of $2\pi$ at $\psi=\pi$.
It can be defined piecewise as
\begin{equation}\label{psi'}
    \psi'=
        \begin{cases}
    \psi & \text{at } \big[0\ ;\ \pi-\epsilon \big)\times S^1,\\
    \psi-2\pi & \text{at }\big(\pi+\epsilon\ ;\ 2\pi\big]\times S^1.
    \end{cases}
\end{equation}
Using $\psi'$ we rewrite $  \tilde{A}^S_{(2)}$ to
\begin{equation}
    \tilde{A}^S_{(2)}=\chi_C \frac{\psi'}{2\pi}\dd\phi.
\end{equation}
which continuously covers the points such that $\psi=0$, i.e. $\psi'=\pi$.

The G-curvature
\begin{equation}
    \tilde{F}:= \dd\tilde{A}^S_{(2)}=\chi_C \frac{1}{2\pi}\dd\psi \wedge \dd\phi = \chi_C \frac{1}{2\pi}\dd\psi' \wedge \dd\phi,
\end{equation}
together with the integral (justifying our choice of the name for $\chi_C$)
\begin{equation}
  \int_{S^1\times S^1} \tilde{F} = 2\pi \chi_C 
\end{equation}
exhausts all the possible values as $\chi_C$ ranges $\mathbb{Z}$.  
Combined with the result of the general theory of principal bundles about the equivalence for the classes of isomorphism of principal $U(1)-$bundles over a manifold $M$ \cite{cohenBunles}
\begin{equation}
    \text{Prin}_{U(1)}(M) \cong H^2(M;\mathbb{Z}),
\end{equation}
where $H^2(M;\mathbb{Z})$ is the second cohomology group of $M$ with coefficients in group $\mathbb{Z}$.
For any compact and orientable Riemann surface, including torus and all surfaces considered in the subsequent sections, this group is isomorphic to $\mathbb{Z}$ which reassures us that we have not missed above any $U(1)$-PFB.

A general connection $A^S_{(i)}, i=1,2$ such that 
\begin{equation*}
    *\dd A^S_{(i)} = \rm const 
\end{equation*}
is given by (\ref{AS}) and an arbitrary closed $1$-form $a$, namely
\begin{equation}
    \tilde{A}^S_{(1)}=\chi_C \frac{\psi}{2\pi}\dd\phi + \alpha \dd\psi + \beta \dd\phi, \ \ \ \tilde{A}^S_{(2)}=\chi_C \frac{\psi'}{2\pi}\dd\phi + \alpha \dd\psi' + \beta \dd\phi', 
\end{equation}
where due to the global gauge transformations the constants $\alpha, \beta$ can be restricted to the interval $[0,1)$.  

The horizon-bundle $H$ is obtained by gluing $S^1\times V_1$ endowed with coordinates $(\tau, \psi, \phi)$ with   $S^1\times V_2$ equipped with coordinates $(\tau', \psi', \phi')$, using the relation (\ref{psi'}) and the following relation  between the remaining  unprimed and primed coordinates, valid on $S^1\times (V_1\cap V_2)$, 
\begin{equation}\label{eq:glueing}
\phi=\phi', \ \      \tau = \tau' -   
    \begin{cases}
    0 & \text{at } \big(\epsilon\ ;\ \pi-\epsilon \big)\times S^1\\
     \chi_C \phi' & \text{at }\big(\pi+\epsilon\ ;\ 2\pi-\epsilon \big)\times S^1, \ \ m\in \mathbb{Z}.
    \end{cases}
\end{equation}
The  null vector field $\ell$ on $H$ is
\begin{equation}\label{ell} 
    \ell =     \begin{cases}
    \partial_\tau & \text{at } S^1\times V_1\\
 \partial_{\tau'} & \text{at }  S^1\times V_2,
    \end{cases}
\end{equation}
while the degenerate metric tensor on $H$ reads
\begin{equation}\label{g}
    g =  \begin{cases}
   \frac{1}{P^2_0}(a^2 \dd \phi^2+2ab\ \dd \psi \dd \phi+ (1+b^2)\dd \psi^2)  & \text{at } S^1\times V_1,\\
 \frac{1}{P^2_0}(a^2 \dd \phi'^2+2ab\ \dd \psi' \dd \phi'+ (1+b^2)\dd \psi'^2). & \text{at }  S^1\times V_2.
 \end{cases}
\end{equation}
Finally,  the  rotation $1$-form potential on $H$ is
\begin{equation}
\label{eq:omega-torus-abstract-horizon-solution}
    \omega =  \begin{cases}
   \kappa \left( \dd\tau + \frac{\chi_C}{2\pi}\psi \dd\phi + \alpha \dd\psi + \beta \dd\phi\right),  & \text{at }  S^1\times V_1;\\
\kappa \left( \dd\tau' + \frac{\chi_C}{2\pi}\psi' \dd\phi' + \alpha \dd\psi' + \beta \dd\phi'\right), & \text{at }  S^1\times V_2.
 \end{cases}
\end{equation}
with the corresponding rotation pseudoscalar
\begin{equation}
    {\Omega}= {\kappa P_0^2}\frac{\chi_C}{2\pi a}.
\end{equation}

\subsubsection{Intermediate summary}
A general vacuum Petrov type D isolated horizon $(H, g, \ell, \nabla)$ whose null flow coincide with the $U(1)$ symmetry has the structure of a principal $U(1)$ fiber bundle over  $2$-torus $S_1\times S_1$ is defined as follows: 
\begin{itemize}
    \item the manifold structure of $H$ is given by  considering $U(1)\times V_1$ endowed with the angle coordinate system $(\tau,\psi,\phi)$ and $U(1)\times V_2$ endowed with the angle coordinate system $(\tau',\psi',\phi')$, and glueing them with  (\ref{psi'}, \ref{eq:glueing}),
    \item  the generator $\ell$ of the null flow is given by (\ref{ell}), 
    \item the rotation $1$-form potential $\omega$ is given by (\ref{eq:omega-torus-abstract-horizon-solution}),
    \item the degenerate metric tensor $g$ is given by (\ref{g}),
    \item  a degenerate metric tensor $g$ and a rotation $1$-form potential on $H$ determine the covariant derivative $\nabla$. 
\end{itemize}
The constants   $a > 0,\ b,\ \alpha,\ \beta,\ P_0\not= 0,\ \kappa \in \mathbb{R}\setminus\{0\},\ \chi_C\in \mathbb{Z}$ are arbitrary. 
Their range may be reduced by the residual discrete coordinate transformation discussed above.    
If $\chi_C\not= 0$, then the structure group of the bundle generated by the flow of $\ell$ is nontrivial, and the horizon does not admit global, spacelike sections. 
If $\chi_C=0$, then the bundle is trivial. 

The formulas (\ref{ell}, \ref{g}, \ref{eq:omega-torus-abstract-horizon-solution}) with $\chi_C=0$ define also a general vacuum Petrov type D isolated horizon $(H, [\ell], g, \omega)$ whose null flow has the structure of a principal $\mathbb{R}$ fiber bundle over  $2$-torus $S_1\times S_1$ if we assume that $\tau$ is coordinate on $\mathbb{R}$. 
In that case, rescalling $\tau$ we can set the surface gravity $\kappa$ to take arbitrarily fixed nonzero value.

\subsection{Solutions of type D equation on a  bundle over Riemann surface with genus $\geq 2$}
\label{sec:sols-Petrov-higher-genus}

Consider an isolated horizon  $(H, g, \nabla, [\ell])$ such that the null flow has the structure of a principal fiber bundle
\begin{equation}
     \pi : H\rightarrow S, \ \ \ G=\mathbb{R}, \ U(1),
\end{equation}
over a Riemann surface $S$ of genus $g\ge 2$. 
The metric tensor $g^{S}$ defined on $S$  has the Ricci tensor 
\begin{equation}
R^{S} = K g^{S},    
\end{equation}
where due to the type D equation $K=\rm const$ \cite{TypeDGenus}, and as such it can be determined by the total area of $S$ and its genus, namely
\begin{equation}
\label{eq:K-higher-genus}
    K=\frac{4\pi (1-\text{genus})}{\text{Area}(g^S)}.
\end{equation}
The degenerate metric tensor on $H$ is the pullback of $g^{S}$
 \begin{equation}\label{gH-g>0}
    g=\pi^*g^S.
\end{equation}

\subsubsection{The trivial bundle case}
\label{sec:trivial-higher-horizons}
As it was pointed out above, in the trivial PFB case  Eq (\ref{Chern}) and the second equality in (\ref{KOmegaconst}) imply 
\begin{equation}
    \Omega = 0,
\end{equation}
hence the  corresponding $G$-connections  (\ref{G-connection}) are flat.
In a consequence, to ensure that (\ref{eq:fPetrov-nonvanishing}) holds it is also required that $K\not=\frac{\Lambda}{3}.$
The horizon over $S$ has the product topology $H = G\times S$ consistent with  $\pi: H\rightarrow S$.
Let us fix a specific product structure. 
Possible groups are $G=\mathbb{R}, \ U(1)$. 
Let $\tau$ be a coordinate along $G$ extended naturally to $H$ using the product structure and such that 
 \begin{equation}
     \ell(\tau) = 1.
 \end{equation}
Then a flat connection on $H$ can be written in terms of a closed 1-form $A^S$ on $S$,  
\begin{equation}
 A = \left( \dd\tau + \pi^*A^S\right) \otimes \partial_\tau,  
\end{equation}
while the formula for the corresponding rotation $1$-form potential such that (\ref{Omega0}) reads
\begin{equation}
\label{omega-triv}
 \omega = \kappa \left( d\tau + \pi^*A^S\right).\end{equation}

Now if the gauge group $G=\mathbb{R}$ then, as explained above, using the scaling freedom of the vector field $\ell$ that defines the isolated horizon structure we can fix $\kappa=1$. 
The automorphisms of the bundle (choice of the Cartesian product) are given by transformations
\begin{equation}
\tau = \tau' + \pi^*f, \ \ \    f\in C^\infty(S).    
\end{equation}

On the other hand, if the gauge group $G=U(1)$, then we can fix the scaling of $\ell$ such that the corresponding coordinate on $U(1)$ ranges from $0$ to $2\pi$.  
Hence a value of $\kappa$ in (\ref{omega-triv}) becomes meaningful. 
However now, the automorphisms of the bundle are
\begin{equation}
e^{i\tau} = e^{i\tau'+i\pi^*f}, \ \ \    \ell(f) = 0,     
\end{equation}
where the smoothness (or suitable differentiability class) is required of the exponentiated function.

\subsubsection{The nontrivial bundle case}
In this case 
\begin{equation}
    G=U(1),
\end{equation}
and the nonequivalent bundles over Riemann surfaces are numbered by the Chern invariant $\chi_C \in \mathbb{Z}$.
The rotation $1$-form potential is given by
\begin{equation}
    \omega = \kappa A
\end{equation}
where $\kappa$ is the surface gravity, and $A$ is a  $1$-form on $H$ such that $A\otimes \ell^*$ is a connection $1$-form of the PFB considered thereon, and   
{\color{black}
 \begin{equation}
 \dd A =  \tilde \Omega \pi^* \eta^S, \ \ \ \ \ \tilde\Omega = \rm const,
\end{equation}
where  $\eta^S$ stands for the area $2$-form  of the metric tensor $g^S$.  
If we find a 1-form $A$ satisfying the above condition for any $\tilde\Omega=\text{const}$ then, as was explained above, this constant necessarily satisfies
\begin{equation}
  \tilde\Omega =  \frac{2\pi \chi_C}{\text{Area} (g^S)} =  \frac{K \chi_C}{2(1-g)}, 
\end{equation} 
where Area$(g^S)$ is the area $S$ according to the metric tensor $g^S$, and  $K$ is the constant Gauss curvature of $g^S$. }
Given a bundle, if we fix a specific  $A$
then every other $A'$ can be written as
\begin{equation}
    A' = A+\pi^* a, \ \ \ \  \dd a=0.
\end{equation}
The gauge equivalence classes $[a]$ are defined by the gauge transformations
$$a \mapsto a -i g^{-1}\dd g, \ \ \ \  g\in C^\infty(S,U(1))$$
as explained in detail in the case of the torus in Sec. 2.3.1. Finally, the rotation pseudo-scalar $\Omega$ is 
\begin{equation}
\label{eq:Omega-higher-genus}
    \Omega=\frac{2\pi \chi_C \kappa}{\text{Area}(g^S)}  = \frac{K\kappa \chi_C  }{2(1-g)}.
\end{equation} 
If $m\not=0$ and since $\kappa\not=0$, also we have that $\Omega\not=0$.
Hence the Gauss curvature $K$ has no forbidden values.  

\section{Embeddability of the isolated horizons with genus $\geq 1$}

In this section, we seek to find the embedding spacetimes for the vacuum Petrov type D isolated horizons $(H, g, \nabla, [\ell])$ constructed in the first part of this work. 
Our approach is to assume that the embedding spacetime admits a $U(1)$-PFB structure, specifically that it has the topology $\mathbb{R}\times H$, and then solve the Einstein equations in terms of elements manifestly compatible with the PFB structure, that is globally defined on $H$.
We solve the Einstein equations for a general, generically nontrivial case, while the trivial limit is possible by assuming that a certain topological charge is zero.
The obtained solution contains as a limit the embedding spacetimes for the trivial bundle horizons. 
In the last part of this section, we exhibit the construction of toroidal embedding spacetimes, where an explicit use of the coordinates on a torus is possible.
{\color{black}The horizons of spherical topology with a nontrivial U(1)-bundle structure have been already found to be embeddable in the accelerated Kerr-NUT-(anti-) de Sitter spacetimes in the generic case, and Taub-NUT-(anti-) de Sitter spacetimes for constant Gaussian curvature  $K$ and rotation invariant $\Omega$ \cite{LDROTransversal}. 
However, because the calculations presented in this section naturally generalize to spherical horizons we include this case.}

In our global construction of the vacuum spacetimes containing Petrov type D horizons we will use the following data:
\begin{itemize}
    \item A $2$ dimensional compact and orientable Riemann surface $S$ without boundary of genus $p$.
    \item A positive definite metric tensor $g^\epsilon$ on $S$, of  constant Gaussian curvature normalized to $\epsilon=0,\pm1$
    \item a $U(1)-$principal fiber  bundle  $\pi:  H \xrightarrow{} S$,
    \item A connection 1-form $A$ on $H$ satisfying the constant curvature condition,compare with (\ref{eq:def-Omega}),
    \begin{equation}
    \label{eq:bundle-structure-eq}
        \dd A =  \tilde\Omega \pi^* \eta^\epsilon, \ \ \ \ \ \ \tilde \Omega = {\rm const}
    \end{equation}
    where $\eta^\epsilon$ is the $2$-area tensor induced by $g^\epsilon$ on $S$. 
\end{itemize}

Actually, the possible values of $\tilde \Omega$ are constrained by the topological invariants if the genus of $S$ is $>1$, namely 
\begin{equation*}
    \tilde\Omega:=\frac{2\pi \chi_C }{\text{Area}(g^\epsilon)} = \frac{\epsilon \chi_C }{2(1-\text{genus})}.
\end{equation*}
 
Recall from the previous sections that a family of the vacuum Petrov type D  isolated horizons is associated with these data. They are given by the following family of pairs $(g,\omega)$, degenerate metric tensor and rotation $1$-form potential, namely 
{\color{black}
\begin{equation}
\label{horizons}
  g = \frac{1}{k^2}\pi^*g^{\epsilon}, \ \ \ \ \omega = \kappa A, \ \ \ \ k,\kappa = \rm const \not=0.    
\end{equation}}
{\color{black}
If the genus is not equal to zero, then the scaling is already determined by the Gauss-Bonnect theorem.
In this case, $k$ is simply the Gaussian curvature of the horizon.
For torus, the area is not determined and $k$ remains arbitrary.}

Given the above ingredients we can build the following spacetime
\begin{equation}
    M = \mathbb{R}\times H,
\end{equation}
and endow it with a metric tensor

\begin{equation}
\label{eq:genTB-metric-ansatz}
    g=-h(r) A^2+\frac{\dd r^2}{f(r)}+R(r)g^\epsilon,
\end{equation}
where $r$ is a coordinate on $\mathbb{R}$, and the functions  $f(r)$, $h(r)$ and $R(r)$ are arbitrary and to be determined by solving Einstein's equations. 
With slight abuse of the notation, we will identify objects defined on the bundle $H$ with their extension to the spacetime. 

Now the goals are as follows
\begin{itemize}
    \item Impose the vacuum Einstein equations with cosmological constant to solve for the functions $f(r)$, $h(r)$ and $R(r)$.
    \item Construct an Eddington-Finkelstein extension of the metric (\ref{eq:genTB-metric-ansatz}) covering possible Killing horizons.
    \item Show that for every Killing horizon derived in the previous section) the degenerate metric tensor $g$  and the rotation  $1$-form potential $\omega$, both defined on $H$, have the form (\ref{horizons}) and that the constants $k,\kappa$ take all possible nonzero values.
\end{itemize}

It is useful to introduce the orthonormal coframe $(e_\epsilon^A)$, $A\in2,3$ on $S$.
The structure equations for $g^\epsilon$ depend only on the connection component $\Gamma^{2}_\epsilon{}_3$.
They read
\begin{equation}
\begin{split}
    &\dd e^2_\epsilon+\Gamma_\epsilon^{2}{}_3\wedge e_\epsilon^3=0,\\
    &\dd e^3_\epsilon-\Gamma_\epsilon^{2}{}_3\wedge e_\epsilon^2=0.
\end{split}
\end{equation}
and are enough to write the only, up to symmetries, component of the two-dimensional curvature 2-form $ R_\epsilon^{A}{}_B$:
\begin{equation}
     R_\epsilon^{2}{}_3=\dd \Gamma^{2}_\epsilon{}_3= R^{2}_\epsilon{}_{323}\ e^{2}_\epsilon \wedge e^{3}_\epsilon= \epsilon\ e^{2}_\epsilon \wedge e^{3}_\epsilon
\end{equation}
To solve the spacetime Einstein equations we employ the orthonormal coframe corresponding to $g$
\begin{equation}
    \begin{split}
        &e^0=\sqrt{h} A,\\
        &e^1=\frac{\dd r}{\sqrt{f}},\\
        &e^2=\sqrt{R}\ e^{2}_\epsilon,\\
        &e^3=\sqrt{R}\ e^{3}_\epsilon.
    \end{split}
\end{equation}
Then the definitions of the objects on the bundle over $S$ allow us to write the structure equations
\begin{equation}
    \begin{split}
        &\dd e^0=-\frac{h'\sqrt{f}}{2 h} e^0\wedge e^1+\sqrt{h}\frac{N\tilde\Omega}{R} e^2 \wedge e^3,\\
        &\dd e^1=0,\\
        &\dd e^2=\frac{R' \sqrt{f}}{2R} e^1\wedge e^2-\Gamma^{2}_\epsilon{}_3\wedge e^3\\
        &\dd e^3=\frac{R' \sqrt{f}}{2R} e^1\wedge e^3+\Gamma^{2}_\epsilon{}_3\wedge e^2.
    \end{split}
\end{equation}
where by $'$ we denote the derivative with respect to $r$.
The detailed computations of spacetimes connection and curvature may be found in the Appendix.

To solve for $f$ and $g$ we add $G_{00}$ and $G_{11}$
\begin{equation}
    G_{00}+G_{11}=\Lambda-\Lambda=0=\frac{\tilde\Omega ^2 h(r)}{2 r^4}-\frac{f'(r)}{r}+\frac{f(r) h'(r)}{r h(r)}.
\end{equation}
Integrating the above we get the relation between $f(r)$ and $h(r)$
\begin{equation}
\label{eq:ratio-f-h}
    \frac{h(r)}{f(r)}=\frac{4 r^2 c_1}{ r^2-\tilde{\Omega}^2c_1},
\end{equation}
where $c_1$ is an integration constant.
The above motivates a following coordinate change.
\begin{equation}
    \bar r=\sqrt{r^2+l^2},
\end{equation}
where the integration constant has been reparametrized $l^2:=-\tilde \Omega^2 c_1$ for $\tilde \Omega c_1<0$.

The metric tensor using the new coordinate reads
\begin{equation}
\label{eq:ansatz-metric-bar}
    g=-h A^2+\frac{\dd\bar r^2}{\bar f}+(\bar r^2+l^2)g^\epsilon,
\end{equation}
where $\bar f:=f \left(\frac{\dd \bar r}{\dd r}\right)^2=f \frac{\bar r^2}{ r^2}$.
After the transformation we have
\begin{equation}
    \partial_{\bar r}\left(\frac{\bar f}{ h}\right)=\frac{\dd r}{\dd \bar r}\partial_r\left(\frac{f}{h}\frac{\bar r^2}{r^2}\right)=0.
\end{equation}
It follows that $h$ and $\bar f$ are proportional $h=:N \bar f$ where $N$ is a real constant.
Note that the apparent singularity at $r=\tilde\Omega^2 c_1$ of (\ref{eq:ratio-f-h}) appears only for the nontrivial bundles, if $m=0$ there is no need to introduce an additional parameter $l^2$ and the role of proportionality constant between $h$ and $f$ is played by the integration constant $c_1$.
The form of the metric tensor given by (\ref{eq:ansatz-metric-bar}) is therefore as general as the one given by (\ref{eq:genTB-metric-ansatz}).
In the subsequent, we drop the bar above $r$ and $h$ and proceed to solve the Einstein equations for the metric tenor
\begin{equation}
\label{eq:generic-TB-bundle-solved}
    g=-f(r) N A^2+\frac{\dd r^2}{f(r)}+(r^2+l^2)g^\epsilon,
\end{equation}
where $l$ is an arbitrary parameter. 
Because the above metric is formally the same as previously, only with $f$ replaced by $N h$ and specified $R$, we reuse the calculations.
Again we have 
\begin{equation}
     G_{00}+G_{11}=0=\frac{f(r) \left(N \tilde\Omega ^2-4 l^2\right)}{2 \left(l^2+r^2\right)^2},
\end{equation}
which establishes 
\begin{equation}
\label{eq:N-solve}
    N=\frac{4l^2}{\tilde\Omega^2}.
\end{equation}
Inserting the above value into $G_{00}$ and solving $G_{00}=-\Lambda$ we get
\begin{equation}
\label{eq:f-solved}
    f(r)=\frac{\epsilon \left(r^2-l^2\right)+ c_1 r - \Lambda\left(\tfrac{1}{3}r^4+2l^2r^2-l^4\right)}{\left(r^2+l^2\right)}.
\end{equation}
with the resulting metric reading
\begin{equation}
    g=-\left(\frac{l \text{Area}(g^\epsilon)}{\pi \chi_C}\right)^2f(r) A^2+\frac{\dd r^2}{f(r)}+(r^2+l^2)g^\epsilon.
\end{equation}
Note that for the toroidal case the function $f(r)$ is already the same as of the toroidal Taub-NUT-(anti-) de Sitter case if one renames $c_1=-2M$.
The remaining freedom is the choice of metric on the torus $S$.

The extension through the horizon requires introducing an (ingoing) Eddington-Finkelstein coframe transformation
\begin{equation}
    \left(\frac{l \text{Area}(g^S)}{\pi \chi_C}\right)A':=\left(\frac{l \text{Area}(g^S)}{\pi \chi_C}\right)A+\frac{\dd r}{f(r)}.
\end{equation}
or equivalently
\begin{equation}
    e'^0:=e^0+e^1.
\end{equation}
Similarly to the torus case, introducing $A$ preserves the bundle structure, equivalently we could have started with an ansatz in the Eddington-Finkelstein form
\begin{equation}
\label{eq:genTB-metric-ansatz-EF}
    g=-\left(\frac{l \text{Area}(g^S)}{\pi \chi_C}\right)^2f(r) A'^2+2\left(\frac{l \text{Area}(g^S)}{\pi \chi_C}\right) A'\dd r+(r^2+l^2)g^S.
\end{equation}
{\color{black}
If the Gaussian curvature of $S$ is nonvanishing the area is already determined by the normalization of $|\epsilon|=1$.
Then in all expressions that will follow one could replace
\begin{equation}
    \frac{l \text{Area}(g^S)}{\pi \chi_C}=\frac{2l |1-p|}{\chi_C}.
\end{equation}
}
Note that $\ell$ is a Killing vector field of $\tilde\Pi^*g^S$ and therefore may be naturally extended to a Killing vector field of $g$ generating a horizon at the hypersurface $r=r_H$, where $f(r_H)=0$.

Now we calculate $K$ and $\Omega$ of the horizon with a metric tensor on section $S$
\begin{equation}
    \prescript{(2)}{}{g_H}=(r^2+l^2)g^\epsilon.
\end{equation}
By assumption $(S,g^\epsilon)$ has the Gaussian curvature $\epsilon$, giving the Gaussian curvature $K$ of the horizon as
\begin{equation}
    K=\frac{\epsilon}{r_H^2+l^2}
\end{equation}
The horizon rotation 1-form $\omega$ and the surface gravity $\kappa$ read
\begin{equation}
\begin{split}
&\omega\otimes\ell:=\nabla \ell \eval_H=\left(\frac{l \text{Area}(g^S)}{\pi \chi_C}\right)\frac{f'}{2}\eval_H e'^0 \otimes e_0=\left(\frac{l \text{Area}(g^S)}{\pi \chi_C}\right)\frac{f'}{2}\eval_HA\otimes\ell,\\
&\kappa:=\nabla_\ell \ell \eval_H =\left(\frac{l \text{Area}(g^S)}{\pi \chi_C}\right)\frac{f'}{2}\eval_H,\\
&\omega=\kappa A.
    \end{split}
\end{equation}
The rotation invariant $\Omega$ associated with the horizon in generalized Taub-NUT-(anti-) de Sitter can be calculated using
\begin{equation}
    \dd \omega =\Omega \eta_H,
\end{equation}
where $\eta_H$ is the volume form associated with $g_H$, as
\begin{equation}
    \label{eq:Omega-embeddedd-all-genus}
    \Omega=\frac{\kappa}{r_H^2+l^2}\frac{2\pi \chi_C}{\text{Area}(g^S)}=\frac{2\pi \chi_C \kappa}{\text{Area}(\prescript{(2)}{}{g_H})}=\frac{\epsilon-(r_H^2+l^2)\Lambda}{r_H(r^2_H+l^2)}
\end{equation}
which exactly reproduces the abstract solution (\ref{eq:Omega-higher-genus}).

Having constructed the toroidal Taub-NUT-(anti-) de Sitter horizon with a structure of a nontrivial $U(1)-$principal bundle over torus is it natural to ask if the bundle structure extends to the bulk spacetimes.

Take the group action generated by the flow of the same Killing vector field $\ell$, now acting on the whole spacetime instead of only the horizon, then the connection 1-form may be defined as
\begin{equation}
\label{eq:omega-spacetime}
A_{\text{ST}}:=\frac{g(\ell,\cdot )}{g(\ell,\ell)}.
\end{equation}
To see that the above indeed is a real-valued part of the connection 1-form one has to check the conditions
\begin{equation}
    A_{\text{ST}}(\ell)=1, \quad \mathcal{L}_\ell A_{\text{ST}}=0,
\end{equation}
which follow immediately from $\ell$ being a Killing vector of the spacetime.

The PFB structure of $\pi$ used in the construction of the spacetime extends naturally to the PFB structure $\Pi$ of the entire spacetime by extension of the real line associated with the coordinate $r$.
\begin{equation}
    U(1) \hookrightarrow P\times \mathbb{R} \xrightarrow{\Pi} S\times \mathbb{R}.
\end{equation}
The connection 1-form defined as (\ref{eq:omega-spacetime}) is simply
\begin{equation}
    A_{\text{ST}}=
    A'-\left(\frac{\pi \chi_C}{l \text{Area}(g^S)}\right)\frac{\dd r}{f(r)}=A
\end{equation}

Finally we note that $A$ (and consequently $A'$, $\omega$ and $A_{\text{ST}}$) for a given $\Omega^S$ is defined uniquely only up to 1-forms $\alpha_1,\alpha_2,\dots,\alpha_{2p}$ generating first de Rham cohomology group $H^1(S,\mathbb{R})\cong \mathbb{R}^{2g}$, where $p$ is the genus of $S$.
In this sense also the constructed spacetime (\ref{eq:genTB-metric-ansatz}) is also unique only up to the $\alpha$'s.

Note that the metric tensors derived above are locally isometric to the generalized Taub-NUT-(anti-) de Sitter (or in the trivial case to the generalized Schwarzschild-(anti-) de Sitter) given by
\cite{griffiths_podolsky_2009}
\begin{equation}
\label{eq:generic-TN-metric}
    g=-f(r)  \bigg(\dd t +l \frac{\ii(\zeta\dd\bar\zeta-\bar{\zeta}\dd\zeta)}{1+\tfrac{1}{2}\epsilon\zeta\bar{\zeta}}\bigg)^2 +f(r)^{-1}\dd r^2 +\left(r^2+l^2\right)\frac{2\dd \zeta \dd \bar{\zeta}}{(1+\tfrac{1}{2}\epsilon\zeta\bar{\zeta})^2},
\end{equation}
where
\begin{equation}
\label{eq:f-fun-generic}
    f(r)=\frac{\epsilon(r^2-l^2)-2M r-\Lambda(\tfrac{1}{3}r^4+2l^2r^2-l^4)}{r^2+l^2},
\end{equation}
and the NUT and mass parameters are denoted by $l$ and $M$, respectively, and $\epsilon=0,\ \pm1$ so that the last term in the above is a flat metric tensor defined on a surface of constant Gaussian curvature equal to $\epsilon$.

The difference between (\ref{eq:generic-TN-metric}) and (\ref{eq:generic-TB-bundle-solved}) is only in the compactification of the surfaces parameterized by $(\zeta,\bar\zeta)$.
Indeed, all tori may be constructed by a suitable identification of sides of a parallelogram, i.e. a quotient $\mathbb{R}^2/\mathbb{Z}^2$.
Similarly, higher genus Riemann surfaces may be constructed as the quotient of a hyperbolic plane by a discrete subgroup $\Gamma\subset \text{PSL}(2,\mathbb{R})$ generated by $2p$ elements $a_1,b_1,\dots,a_p,b_p$ satisfying single defining relation
\begin{equation}
    a_1 b_1 a_1^{-1} b_1^{-1}\dots a_g b_p a_p^{-1} b_p^{-1}=1.
\end{equation}
Then this surface is a hyperbolic $4p$-gon  with a pair of edges identified.

\subsection{Toroidal Taub-NUT-(anti-) de Sitter and nontrivial bundles}
Although the toroidal case is covered by the construction in the previous section, there is an advantage to consider it separately.
Namely on a torus global coordinates may be easily introduced which allows for an explicit construction of both the connection $A$ and the metric on the base space $g^S$.

Consider the metric tensor (\ref{eq:generic-TB-bundle-solved}) with $f(r)$ and $N$ given by (\ref{eq:f-solved}) and (\ref{eq:N-solve}) and $\epsilon=0$. The most general metric $g^\epsilon$ may be given for torus as
\begin{equation*}
    g^\epsilon=(a^2 \dd \phi^2+ 2 a b\ \dd \phi \dd \psi+ (1 + b^2)\dd \psi^2 )
\end{equation*}
which implies that
\begin{equation}
    A=\dd \tau +\frac{\chi_C}{2\pi}\psi\dd\phi,
    % +\alpha\dd\psi+\beta\dd\phi,
\end{equation}
where $\tau\in [0;2\pi)$ and $\partial_\tau$ generates the Killing horizon.

After introducing Finkelstein-Eddington coordinates it is easy to see that the nondegenerate metric tensor on a toroidal section of the horizon is
\begin{equation}
    \prescript{(2)}{}g_H=(r_H^2+l^2)g^\epsilon,
\end{equation}
which compared with the isolated horizon given by the solution to the Petrov type D equation yields $r_H^2+l^2=P_0^{-2}$.
Comparing the rotation 1-form $\omega=\kappa A$ for the above horizon with the solution to the Petrov type D equation given by (\ref{eq:omega-torus-abstract-horizon-solution}) it seems that only the horizons with $\alpha=0=\beta$ are embeddable.
To embed the general class it is sufficient to introduce a local coordinate transformation 
\begin{equation}
\label{eq:v-to-vAB}
    \tau \mapsto \tau + \alpha\psi + \beta\phi.
\end{equation}
Importantly the above is only a local diffeomorphism.
It does not extend to global diffeomorphism as the transformation is not well defined whenever the coordinates $(\psi,\phi)$ are not well defined, namely if $\psi=0$ or $\phi=0$.
Knowing the first de Rham cohomology group of the torus $H^1_{\text{dR}}\cong \mathbb{R}^2$ we conclude that the above transformation accounts for all solutions to (\ref{eq:bundle-structure-eq})).
Finally, the rotation invariant of the horizon reads
\begin{equation}
    \Omega=\frac{\chi_C\kappa}{2\pi (r^2_H+l^2)}=\frac{ l \Lambda}{r_H}.
\end{equation}

It should be noted that the formula for $\Omega$ (\ref{eq:Omega-embeddedd-all-genus}) does not have limit to $\Lambda=0$ in the toroidal case.
Then we have
\begin{equation}
    f(r)=\frac{-2M r}{r^2+l^2},\quad r_H=0,
\end{equation}
which results
\begin{equation}
    \Omega=-\frac{2 M}{l^3}.
\end{equation}

\subsection{Generalized Schwarzschild-(anti-) de Sitter  and trivial horizons}
\label{sec:Embeddability-trivial}
If the horizon bundle is trivial i.e. $\chi_C=0=\Omega^S$ then the solutions (\ref{eq:genTB-metric-ansatz-EF}) obtained in the previous section do not have a well defined limit.
Instead, we go back to the derivation earlier in that section.
Due to the triviality the structure equation containing $\dd A =0$ changes.
Now the choice of $R(r)$ giving the proportionality between $f(r)$ and $h(r)$ is standard in the Schwarzschild case, i.e. $R(r)=r^2$.
$G_{00}+G_{11}=0$ gives
\begin{equation}
    h(r)=c_1 f(r).
\end{equation}
Solving $G_{00}=-\Lambda$ we recover
\begin{equation}
    f(r)=\epsilon+\frac{c_2}{r}-\frac{\Lambda  r^2}{3},
\end{equation}
recovering the generalized Schwarzschild-(anti-) de Sitter for $c_2=-2M$
with the metric tensor reading
\begin{equation}
    g=-c_1 f(r) A^2+\frac{\dd r^2}{f(r)}+r^2 g^\epsilon.
\end{equation}

{\color{black}
If the fiber is $\mathbb{R}$ then $c_1$ is arbitrary and may be set to $1$, similarly to the arbitrary choice rescaling of the time coordinate in Schwarzschild spacetime.
Otherwise is the fiber is $U(1)$ and $c_1$ is determined by the range of the fibre coordinate.}

\section{Summary}
We considered isolated horizons whose null flow has the structure of a  PFB over a Riemann surface.
The case of our particular interest was a nontrivial bundle and genus higher than $0$, however, the trivial bundle case also appears as a special, degenerate sector. 
First, we considered unembedded isolated horizons, that is certain $3$-surfaces endowed with isolated horizon structure.  
For them, we solved the vacuum Petrov type D equation with (possibly zero) cosmological constant (Sect. 2). 
In the case of $2$-torus we used explicit coordinates (Sec \ref{sec:torus-horizons}). 
In the case of higher genus, our considerations were coordinate-free (Sec. \ref{sec:sols-Petrov-higher-genus}).  
Next, we constructed a family of vacuum spacetimes containing the horizons. 
The spacetimes are of the Petrov type D, hence the local formula for the metric tensor is that of Demiański-Plebański, however, our emphasis was on the global structure of spacetime that is the cartesian product of the horizon and the real axis representing the $4$th direction, and that it has the symmetries of the horizon. 
Upon those assumptions we have solve the vacuum  Einstein equations and derived a general solution.
Among the general solution the horizons and their embeddings characterized by a trivial bundle structure, either with line or circle fibers, were found.
Such horizons were found to be embeddable in spacetimes locally isometric to generalized Schwarschild-(anti-)de Sitter.

Finally, we note that the constructed horizons are characterized by:
the genus $p$ of the base space $S$ of the horizon, 
the Chern number $\chi_C$ accounting for all possible $U(1)-$ bundles one could construct over $S$, a constant curvature metric tensor on $S$ accounting for all possible ways of compactifing a torus or a hyperboloid and a $U(1)$-connection with constant curvature.
The $U(1)$-connections satisfying the same curvature condition may still differ by a closed but not exact 1-form, leading to a globally different horizon.
In the nontrivial case, the surface gravity $\kappa$ is not free but is instead determined by the above.
Upon embedding $K$ and $\Omega$ are determined by the radius of the horizon $r_H$ and the NUT parameter $l$.
The NUT parameter determines only whether the bundle structure is trivial or not, by itself it does not determine the topological invariant $\chi_C$.

This work closes the classification of type D isolated horizons with a bundle structure generated by the horizon-generating null field.
{\color{black}
The classification may be done in two directions; Firstly, one has to specify the global topology of the horizon, the important part being whether the fibration is trivial or not.
Second, one has to specify the topology of the space of the null generators (for the trivial bundles this is simply a section of the horizon).
The trivial, $S^2\times\mathbb{R}$ horizons and their embeddings into Kerr-(anti-) de Sitter have been found in \cite{localnohairPhysRevD.98.024008}, while nontrivial horizons with spherical space of null geodesics were considered in \cite{hopf}, however, the full embedding into the accelerated Kerr-NUT-(anti-) de Sitter spacetimes was established only recently in \cite{LDROTransversal}.
Obviously, in the trivial case, the fiber may be always changed from $\mathbb{R}$ to $S^1$.
The same is not true for nontrivial horizons whose fibers must be circles.
It turns out the Petrov type D horizons with nonspherical but still compact spaces of null generator its curvature must be constant (vanishing or negative), which was shown in \cite{TypeDGenus} for the trivial bundle case.
This paper occupies the remaining place in the classification, i.e. it covers the higher genus horizons with nontrivial fibration topology.

In \cite{LDROTransversal} authors have also considered spherical horizons with bundle structure generated by a field transversal to the null one.
It remains an open problem whether a similar construction may be done in the higher genus case.
}
A remaining open problem is the constructing of horizons with bundle structure generated by a field transversal to the null one and with base space other than the sphere.
Similarly, it is not known whether spherical isolated horizons without axial symmetry exist.

\noindent{\textit{Acknowledgements}  }
The authors were supported by the Polish NCN grant for the project OPUS 2021/43/B/ST2/02950.

\appendix
\section{Einstein equations for the spacetimes with $U(1)-$PFB structure}
The (Levi-Civita) connection 1-forms $\Gamma^\mu{}_{\nu}$ may be written as
\begin{equation}
\begin{split}
&\dd e^\mu+\Gamma^{\nu}{}_{\alpha} \wedge e^\alpha=0,\\
&\Gamma^0{}_1=\Gamma^1{}_0=\frac{h'\sqrt{f}}{2 h }e^0,\\
&\Gamma^0{}_2=\Gamma^2{}_0=\frac{N\tilde\Omega}{2R} \sqrt{h} e^3,\\
&\Gamma^0{}_2=\Gamma^3{}_0=-\frac{N\tilde\Omega}{2R} \sqrt{h} e^2,\\
&\Gamma^1{}_2=-\Gamma^2{}_1=-\frac{R' \sqrt{f}}{2R}e^2,\\
&\Gamma^1{}_3=-\Gamma^3{}_1=-\frac{R' \sqrt{f}}{2R}e^3,\\
&\Gamma^2{}_3=-\Gamma^3{}_2=\Gamma^2_\epsilon{}_3+\frac{N\tilde\Omega}{2R} \sqrt{h} e^0.
\end{split}
\end{equation}
Similarly for the curvature 2-forms $\Omega^\mu{}_{\nu}$
\begin{equation}
    \begin{split}
        \Omega^\mu{}_\nu&=\dd \Gamma^\mu{}_\nu+\Gamma^\mu{}_\alpha\wedge\Gamma^\alpha{}_\nu,\\
        \Omega^0{}_1&=-A e^0 \wedge  e^1+B e^2 \wedge  e^3,\\
        \Omega^0{}_2&=-C e^0 \wedge  e^2+\frac{1}{2}B e^1 \wedge  e^3,\\
        \Omega^0{}_3&=-C e^0 \wedge  e^3-\frac{1}{2}B e^1 \wedge  e^2,\\
        \Omega^1{}_2&=\frac{1}{2}Be^0 \wedge  e^3-D e^1 \wedge  e^2,\\
        \Omega^1{}_3&=-\frac{1}{2}B e^0 \wedge  e^2-D e^1 \wedge  e^3,\\
        \Omega^2{}_3&=-B e^0 \wedge  e^1+E e^2 \wedge  e^3,
    \end{split}
\end{equation}
where
\begin{equation}
    \begin{split}
        A&=\frac{(2 f h h'' -f \,h'^{2}+f' h h')}{4 h^{2}} ,\\
        B&=\frac{\tilde\Omega  \mathrm{\sqrt{f}}\, \left(h' R-h R'\right)}{2 R^{2} \mathrm{\sqrt{h}}},\\
        C&=\frac{(\tilde\Omega^{2} h^{2}+R R' f h')}{4 R^{2} h},\\
        D&=\frac{(R R' f' +2 R R'' f -f \,R'^{2})}{4 R^{2}},\\
        E&=\frac{\left(3 \tilde\Omega^{2} h-f \,R'^{2}+4 \epsilon R\right)}{4 R^{2}}.
    \end{split}
\end{equation}

By standard conventions we have 
\begin{equation*}
    \Omega^\mu{}_\nu =\tfrac{1}{2} R^\mu{}_{\nu\alpha\beta} e^\alpha \wedge e^\beta,\quad R_{\mu\nu}=R^\alpha{}_{\mu\alpha\nu},\quad \text{Ric}=R^\alpha{}_\alpha,\quad E_{\mu\nu}=R_{\mu\nu}-\tfrac{1}{2} \text{Ric}\ g_{\mu\nu} 
\end{equation*}
and so
\begin{align}
        &R_{00}=A+2C,\quad R_{11}=-A-2D,\quad R_{22}=R_{33}=-C-D+E,\\
    &\text{Ric}=-2A-4C-4D+2E,\\
    &G_{00}=-2D+E,\quad G_{11}=2C-E,\quad E_{22}=E_{33}=A+C+D.
\end{align}

\bibliographystyle{unsrt}
\bibliography{references}

\end{document}